\documentclass[aps, twocolumn, showpacs, preprintnumbers, amsmath, amssymb, prl]{revtex4-1}
\usepackage{amssymb}
\usepackage{mathrsfs}
\usepackage{amsmath}
\usepackage{graphicx}
\usepackage{epstopdf}
\usepackage{float}
\usepackage{multirow}
\usepackage{array}
\usepackage{color}
\usepackage[colorlinks=true, letterpaper=true, pdfstartview=FitV, linkcolor=blue, citecolor=blue, urlcolor=blue]{hyperref}

\makeatletter
\newcommand{\Rmnum}[1]{\expandafter\@slowromancap\romannumeral #1@}
\makeatother

\begin{document}

\title{Theory of Nonlinear Response for Charge and Spin Currents}

\author{Zhi-Fan Zhang$^{1,2}$}

\author{Zhen-Gang Zhu$^{1,2,3}$}
\email{zgzhu@ucas.ac.cn}

\author{Gang Su$^{4}$}
\email{gsu@ucas.ac.cn}

\affiliation{$^{1}$ School of Electronic, Electrical and Communication Engineering, University of Chinese Academy of Sciences, Beijing 100049, China.\\
$^{2}$ School of Physical Sciences, University of Chinese Academy of Sciences, Beijing 100049, China. \\
$^{3}$ CAS Center for Excellence in Topological Quantum Computation, University of Chinese Academy of Sciences, Beijing 100049, China.\\
$^{4}$ Kavli Institute for Theoretical Sciences, University of Chinese Academy of Sciences, Beijing 100190, China.
 }


\begin{abstract}
The nonlinear Hall effect, which is the second-order harmonic charge Hall effect from the Berry curvature dipole in momentum space, has received much attention recently. As the responses to higher harmonics of the driving ac electric field are prominent and measurable, we develop a general nonlinear theory by taking the charge and spin currents as well as the longitudinal and transverse effects into account. We introduce the expansion order of the electric field and Berry curvature multipole moment, where the Berry curvature dipole is a particular one manifesting itself at the second harmonic order  and the second expansion order of the electric field. There are four cases with conserving or breaking the time-reversal symmetry (TRS) and inversion symmetry (IS). We find a specific ``selection rule" that only longitudinal odd harmonic order charge currents exist for conserving both the TRS and IS, and with breaking both symmetries, all harmonic order charge and spin currents are nonzero. With conserving TRS and breaking IS, the charge Hall current exists at even harmonic order, and the longitudinal charge current occurs at odd harmonic order. Only the longitudinal spin current survives at even harmonic order. With breaking TRS and conserving IS, only odd harmonic order  charge and spin currents can appear. Moreover, we observe that every harmonic order current contains a series of infinite-order expansion of the electric field. We further show that the Berry curvature dipole and quadrupole can be determined by measuring the second and fourth harmonic order currents in experiments. This may open a way to explore the higher responses of an ac driving system.
\end{abstract}
\maketitle

\emph{Introduction.}---
The Hall effect is closely related to the topological properties of condensed matter \cite{Klitzing,Prange,Xiao}.
It describes that a transverse charge current can be induced by a longitudinal applied electric field $\mathbf{E}$ in a perpendicular external magnetic field $\mathbf{B}$.
The induced charge current is a linear response to  $\mathbf{E}$ on account of  breaking the time-reversal symmetry (TRS).
For a TRS system, as the Berry curvature is odd in momentum space \cite{Xiao}, $\Omega(\vec{k})=-\Omega(-\vec{k})$, its integral weight by the equilibrium Fermi distribution is forced to vanish. Thus, the linear Hall current disappears because of the limitation of the TRS.
However, it was proposed that a nonlinear charge Hall effect, i.e. the second-order harmonic charge Hall (2HCH) current from the Berry curvature dipole (BCD) in momentum space can exist in the broken inversion symmetry (IS)  \cite{Low2015,Sodemann,Du}, which has been experimentally confirmed \cite{Qiong,Kang}.
Recently, the 2HCH effect has been studied in many materials, such as
Weyl semimetals \cite{Du,Qiong,Kang,Zhang,Chen}, topological insulators \cite{Xu,Facio}, transition metal dichalcogenides family \cite{You,Son}, strained graphene \cite{Battilomo}, ferromagnetic materials with $Tt_{1/2}$ symmetry \cite{Ding} , ferroelectric metals \cite{Wang,Xiao2020b,Wang2019,Kim2019,Xiao2020} and piezoelectric-like device \cite{Xiao2020a}.
Similarly, as the linear response of the spin Hall effect is connected with Berry phase \cite{Ma,Shen}, the nonlinear spin Hall effect, i.e. the second-order harmonic spin Hall effect(2HSH) is also discussed in the transition metal dichalcogenides \cite{Yu}, topological Dirac semimetals \cite{Araki2018}, and  two-dimensional(2D) Rashba-Dresselhaus system \cite{Hamamoto,Pan2019}, etc.

Within the linear response theory, Onsager \cite{Onsager}, Kubo \cite{Kubo}, and Luttinger \cite{Luttinger} explained how to relate the coefficients of the driving fields to microscopic quantities, but  the higher-order terms are less clear.
Since the second order response can be measured by observing the response signal with $2\omega $ frequency ($\omega $ is the frequency of the external field)  by a lock-in amplifier in a phase sensitive way \cite{Qiong}, the other order signals with $n\omega$ frequency can be measured as well.
It is desired to study the higher orders even to the infinite order response for charge and spin transport either in longitudinal or in transverse direction.
On one hand, the infinite-order expansion exposes the general hierarchical structure of response theory, especially by symmetry regulation.
On the other hand, the higher order response shows the sensitivity of nonlinear transport clearly, which is more helpful for us to understand the distribution of Berry curvature in $k$ space (Berry curvature dipoles, quadrupoles, and so on).
Therefore, we present a general formalism for the nonlinear responses of charge and spin current by introducing harmonic order  and the expansion order of the electric field.
Conserving or breaking the TRS or IS, we discuss four cases and demonstrate the conditions for existence or disappearance of the even or odd harmonic order charge and spin currents.
We find every harmonic order contains an infinite expansion order of the electric field  in which the lowest expansion order of the electric field has the same order as that of the harmonic order.
To characterize this, we introduce the Berry curvature multipole moment in which the Berry curvature dipole is just induced for the second harmonic order  and second expansion order of the electric field. We particularly study the role of Berry curvature quadrupole (BCQ) moment in a tilted Dirac model, which can be measured experimentally in terms of the given vector expressions.

\begin{table*}[btp]
\renewcommand\arraystretch{2}
\centering
\caption{A ``selection rule" for charge and spin currents in four symmetry cases.}
\begin{tabular*}{18cm}{@{\extracolsep{\fill}}lccp{6.5cm}p{6.5cm}}
\hline\hline
 Case     & TRS  & IS    & Charge current & Spin current  \\
\hline
I \cite{Sodemann,Du}            &Yes           & No
  &$j_a^{2m\omega } =j_{T_a}^{2m\omega }, \quad j_a^{(2m+ 1)\omega } =j_{L_a}^{(2m + 1)\omega }.$      &$j_{a,{s_v}}^{2m\omega } = \sum\limits_I\int_k{{\bar{ \hat {J}}}_{a,{s_v}}}  \mathcal{F}^{I}_{2m}, \quad j_{a,{s_v}}^{(2m+1)\omega } =0.$\\
\hline
 II                                              & No            &Yes
  &$j_a^{2m\omega}=0, \quad j_a^{(2m+1)\omega}=j_{T_a}^{(2m+1)\omega}+j_{L_a}^{(2m+1)\omega}.$  &$ j_{a,{s_v}}^{2m\omega } = 0, \quad j_{a,{s_v}}^{2m\omega } = \sum\limits_I\int_k{{\bar{ \hat {J}}}_{a,{s_v}}}  \mathcal{F}^{I}_{2m}.$ \\
\hline
III                                                &Yes                   &Yes
  &$j_a^{2m\omega}=0, \quad j_a^{(2m+1)\omega}=j_{L_a}^{(2m+1)\omega}.$    & $j_{a,{s_v}}^{2m\omega }=0, \quad j_{a,{s_v}}^{(2m+1)\omega } =0.$ \\
\hline
  IV                                                        & No   & No
  &$j_a^{m\omega } =j_{T_a}^{m\omega }+j_{L_a}^{m\omega }.$    & $j_{a,{s_v}}^{m\omega}=\sum\limits_I\int_k{{\bar{ \hat {J}}}_{a,{s_v}}}  \mathcal{F}^{I}_{m}. $    \\
\hline\hline
\end{tabular*}
\label{table}
\end{table*}

\emph{Boltzmann distribution function.}---
We suppose that there is only an ac driving electric field $\mathscr{E}_c(t)= {\mathop{\rm Re}\nolimits} \{ {{\cal E} _c}{e^{i\omega t}}\} $, ($c=x, y, z$), that oscillates harmonically with time but is uniform in space.
Based on the existing theoretical studies \cite{Sodemann,Du} and experiments \cite{Qiong,Kang},  we only consider the isotropic model and elastic scattering effect. Thus, the Boltzmann equation under the relaxation-time approximation is applicable \cite{Mahan,Abrikosov1988}
\begin{equation}
- \frac{{e\tau }}{\hbar }{\mathscr{E}_c}{\partial _c}f + \tau {\partial _t}f = {f_0} - f,
\label{infinite order1}
\end{equation}
where $f_0$ is the equilibrium distribution in the absence of $\mathscr{E}_c(t)$, $\tau$ is the relaxation time of electrons, ${\partial _c}= \frac{\partial }{{\partial {k_c}}}$, and ${\partial _t}= \frac{\partial }{{\partial {t}}}$.
We may write the infinite-order distribution function as $f = {\mathop{\rm Re}\nolimits} \{ {f_0} + {f_1} + {f_2} +  \cdots +{f_n}+\cdots\} $.
Generally we get a recurrence structure
$f_n = \sum_{j=0,1,2, \cdots, N} {f_n^{(n - 2j)\omega }{e^{i(n - 2j)\omega t}}}$,
where $N=n/2$ for even $n$, and $N=(n-1)/2$ for odd $n$. Every $f_{n}$ can be analytically derived from the Boltzmann equation. 
We leave the details of derivation in the Supplemental Material \cite{SM}.

\emph{General formulation.}---Driven by an external electric field, the response charge current density is given by
\begin{equation}
\vec{j} =  - \sum\limits_I {e\int_k {{\vec{v}^I}f^I} },
 \label{infinite order2}
\end{equation}
where $I=\pm$ are the band indices, $\int_k = \int{\frac{{{d^2}k}}{{{{(2\pi )}^2}}}} $,  and ${{(v^I)}_a} = v_k^a + {\varepsilon ^{abc}}\Omega _b^I{\dot k_c} = \frac{{\partial \hat{H}}}{{\partial {k_a}}} - \frac{e}{\hbar }{\varepsilon ^{abc}}\Omega _b^I{\mathscr{E}_c}$, $v_k^a$ represents the group velocity of electrons in the direction of $a$, $\varepsilon ^{abc}$ is the Levi-Civita symbol, and the second term of $(v^I)_a$ indicates the anomalous velocity of electrons carried by Berry curvature $\Omega_b^I$ of the $I$-th band, and $f^I$ is the same as the $f$ defined above with the band index $I$.

According to a lengthy calculation in the Supplemental Material \cite{SM}, we emphasize that the harmonic order is not associated with the expansion order of the electric field. Therefore, we focus on distinguishing the harmonic order and the expansion order of the electric field. First, the charge current density $\vec{j}$ can be expanded in the harmonic order as
${j_a} = {\mathop{\rm Re}\nolimits} \{ j_a^0 + j_a^{\omega}{e^{i\omega t}} + j_a^{2\omega }{e^{i2\omega t}}+  \cdots+ j_a^{n\omega}{e^{in\omega t}}+\cdots \}$, and we get the $m$-th harmonic order charge current
\begin{equation}
j_a^{m\omega }= \sum\limits_I \frac{{{e^2}}}{{2\hbar }}\int_k {{\varepsilon ^{abc}}}\Xi_{c} \Omega _b^I   - \sum\limits_I {e\int_k {v_k^a} {\cal F}_m^I},
\label{infinite order3}
\end{equation}
where $\Xi_{c}=\left( {{{\cal E} _c}{\cal F}_{m - 1}^I + {\cal E}_c^*{\cal F}_{m + 1}^I} \right)$, $\mathcal{F}^{I}_{n}=\sum\limits_{l= 0,1,2,\cdots}f^{I,n\omega}_{n+2l}$.
If we define a scalar $\Lambda_n$ and a vector $\vec{V}_n$, i.e. $\Lambda _n= \sum\limits_I {\int_k {\Omega ^I\mathcal{F}_n^I} } $, $\vec{V}_n= \sum\limits_I {\int_k {\vec{v}_k\mathcal{F}_n^I} } $, the m-th harmonic order  charge current  in Eq. (\ref{infinite order3}) will become
\begin{equation}
\vec{j}^{m\omega } = \frac{{{e^2}}}{{2{\hbar }}}\left[ {\hat{z} \times\vec{ {\cal E}}(\Lambda _{m - 1}) + \hat{z}\times\vec{ {\cal E}}^*(\Lambda _{m + 1})} \right] - e{\vec{V}_m}.
\label{infinite order4}
\end{equation}
The first term in square bracket is the $m$-th  harmonics ac Hall current density (i.e., transverse current), 
which we call $\vec{j}_{\text{T}}^{m\omega}$. The second term is the longitudinal current, 
which we refer to as $\vec{j}_{\text{L}}^{m\omega}$.
Next we introduce the expansion order of the electric filed current for the $m$-th harmonic order  current in the form of
\begin{equation}
j_a^{m\omega } = j_a^{m\omega}({\cal E}^m) + j_a^{m\omega}({\cal E}^{m+2}) + j_a^{m\omega}({\cal E}^{m+4}) +  \cdots.
 \label{infinite order5}
\end{equation}
It shows that the $m$-th harmonic order current density contains an infinite order of the electric field (the lowest expansion order of the electric field  is the $m$-th) current density.
 For $m=0$ (the zeroth harmonic order), the first term in Eq. (\ref{infinite order5}) is $j_a^{0}({\cal E}^0) =  - \sum\limits_I {e\int_k {v_k^a} {f_0^I}} $, which is independent of  ${\cal E}$ and should disappear at equilibrium.  However, the other terms contain higher orders of ${\cal E}$. For example, the second term is proportional to ${\cal E}^{2}$, which means a nonlinear dc response (the second expansion order of the electric field) due to the ac driving external electric field
%
and was obtained previously \cite{Sodemann}. One may note that even in this case ($m=0$) there are higher expansion order of the electric field dc charge current up to infinite. 
Equation (\ref{infinite order5})is also valid for ${\cal E}=0$, where all components of current are zero.
For $m=1$ (the first harmonic order), the first expansion order of the electric field is $j_a^{\omega }({\cal E}^1)= \sum\limits_I\left\{\frac{{{e^2}}}{{2\hbar }}\int_k {{\varepsilon ^{abc}}} {\Omega _b^I}{{\cal E}^{*} _c}{f_0^I} - e\int_k {v_k^a} f_1^{I,\omega }\right\} $, which is the conventional result derived from the linear response theory including Hall and drift current. 

The spin current operator can be defined by 
$\hat{J}_{a, s_{v}} \equiv \frac{1}{4}\left\{v_k^a, \sigma_{v}\right\}$
\cite{Hamamoto, Murakami, Murakami2004, Sinova}, where $\hbar=1$ is assumed. The harmonic order spin current is derived as $j_{a,{s_v}} = {\mathop{\rm Re}\nolimits} \{ j_{a,{s_v}}^0 + j_{a,{s_v}}^\omega {e^{i\omega t}} + j_{a,{s_v}}^{2\omega }{e^{i2\omega t}}+\cdots  + j_{a,{s_v}}^{n\omega }{e^{in\omega t}}+\cdots \}$, with the components
\begin{equation}
j_{a,{s_v}}^{m\omega } = \sum\limits_I {\int_k {{{\bar{\hat{J}}}_{a,{s_v}}}{\cal F} _m^I} },
 \label{infinite order6}
\end{equation}
where ${{\bar{ \hat {J}}}_{a,{s_v}}} = \left\langle {I,{\bf{k}}\left| {{{\hat{J}}_{a,{s_v}}}} \right|I,{\bf{k}}} \right\rangle$ , $a$ represents the direction of flow, and
 $v$ indicates the direction of spin polarization.
Substituting $f_m^I$ into Eq. (\ref{infinite order6}), we get 
\begin{equation}
j_{a,{s_v}}^{m\omega} 
= j_{a,{s_v}}^{m\omega}({\cal E}^m)  + j_{a,{s_v}}^{m\omega}({\cal E}^{m+2})   
+\cdots.
 \label{infinite order7}
 \end{equation}
This equation describes the connection between the $m$-th harmonic order spin current density and the expansion order of the electric field spin current density.

\emph{Symmetry analysis.}---
Let us consider the general formula of the charge [Eq. (\ref{infinite order3})] and spin current [q. (\ref{infinite order6})].
We first consider the Hamiltonian with the TRS, satisfying $H(k,\sigma ) = H( - k, - \sigma )$, and every eigenstate has a TRS partner that carries opposite momentum and opposite spin. The Berry curvature $\Omega _b^I$ and the group velocity $ v_k^a$  are thus both odd (${\Omega _b} \to  - {\Omega _b}; v_a^k \to  - v_a^k$). The distribution function $f_n^I  \propto {({{\cal E} _c}{\partial _c})^n}{f_0} $ and  $ \mathcal{F}_n^I$ is even (odd) for even (odd) $n$. The even and odd harmonic order charge current give different results
\begin{eqnarray}
j_a^{2m\omega } =j_{T_{a}}^{2m\omega }, \hspace{0.5 cm} j_a^{(2m + 1)\omega }=j_{L_{a}}^{(2m+1)\omega }.
\label{infinite order8}
\end{eqnarray}
A remarkable conclusion can be drawn from Eq. (\ref{infinite order8}) that only even harmonic order Hall currents and odd harmonic order longitudinal currents exist in the presence of TRS.
The so-called nonlinear Hall current (i.e. the 2HCH current) due to the contribution of the Berry curvature dipole \cite{Sodemann,Du} is just the leading-order non-vanishing term in the total charge Hall current, i.e. the first equation in Eq. (\ref{infinite order8}).
%
Similarly, under the TRS the spin current operator remains intact ($\hat{J}_{a, s_{v}} \to\hat{J}_{a, s_{v}}$), and we get
\begin{equation}
j_{a,{s_v}}^{2m\omega }\neq 0,  
\hspace{0.5 cm}
j_{a,{s_v}}^{(2m+1)\omega } =0.
 \label{infinite order9}
\end{equation}
Comparing Eq. (\ref{infinite order8}) with  Eq. (\ref{infinite order9}), one may uncover that the odd-order distribution function, i.e., $\mathcal{F}_{2\ell + 1}^I$, where $\ell$ is an integer,  only contributes to the charge current; whereas the even-order distribution function, i.e., $\mathcal{F}_{2\ell}^I$,  only contributes to the spin current under the TRS. If we focus on the longitudinal charge and spin current, we find that the disappearance and appearance of the charge and spin current are just opposite, occurring in alternative odd and even harmonic order  ways\cite{Hamamoto}.

We discuss the second case in which the Hamiltonian satisfies $H(k,\sigma ) = H( - k, \sigma )$ under the IS. The Berry curvature $\Omega _b^I$ is conserved (${\Omega _b} \to  {\Omega _b}$), and we have
\begin{eqnarray}
j_a^{(2m+1)\omega} =j_{T_a}^{(2m+1)\omega}+j_{L_a}^{(2m+1)\omega}\neq 0, 
j_a^{2m\omega}=0.
 \label{infinite order10}
\end{eqnarray}
The spin current density can be obtained 
\begin{equation}
j_{a,{s_v}}^{2m\omega } = 0, \hspace{0.6 cm}
j_{a,{s_v}}^{(2m+1)\omega }\neq 0.    
 \label{infinite order11}
\end{equation}
If the TRS and IS exist simultaneously, the Berry curvature and the spin current operator must disappear throughout the  Brillouin zone. Hence there is no Hall phenomenon and only the longitudinal current is left, yielding
\begin{equation}
j_a^{(2m+1)\omega}=j_{L_a}^{(2m+1)\omega}\neq 0.
 \label{infinite order12}
\end{equation}
According to the above analysis, Table \ref{table}
summarizes a ``selection rule" for the charge and spin current in four cases with different symmetries. It can be seen that case I is just the case discussed in Refs. \cite{Sodemann,Du}.
\begin{figure}[tbp]
\centering
\includegraphics[width=1\linewidth]{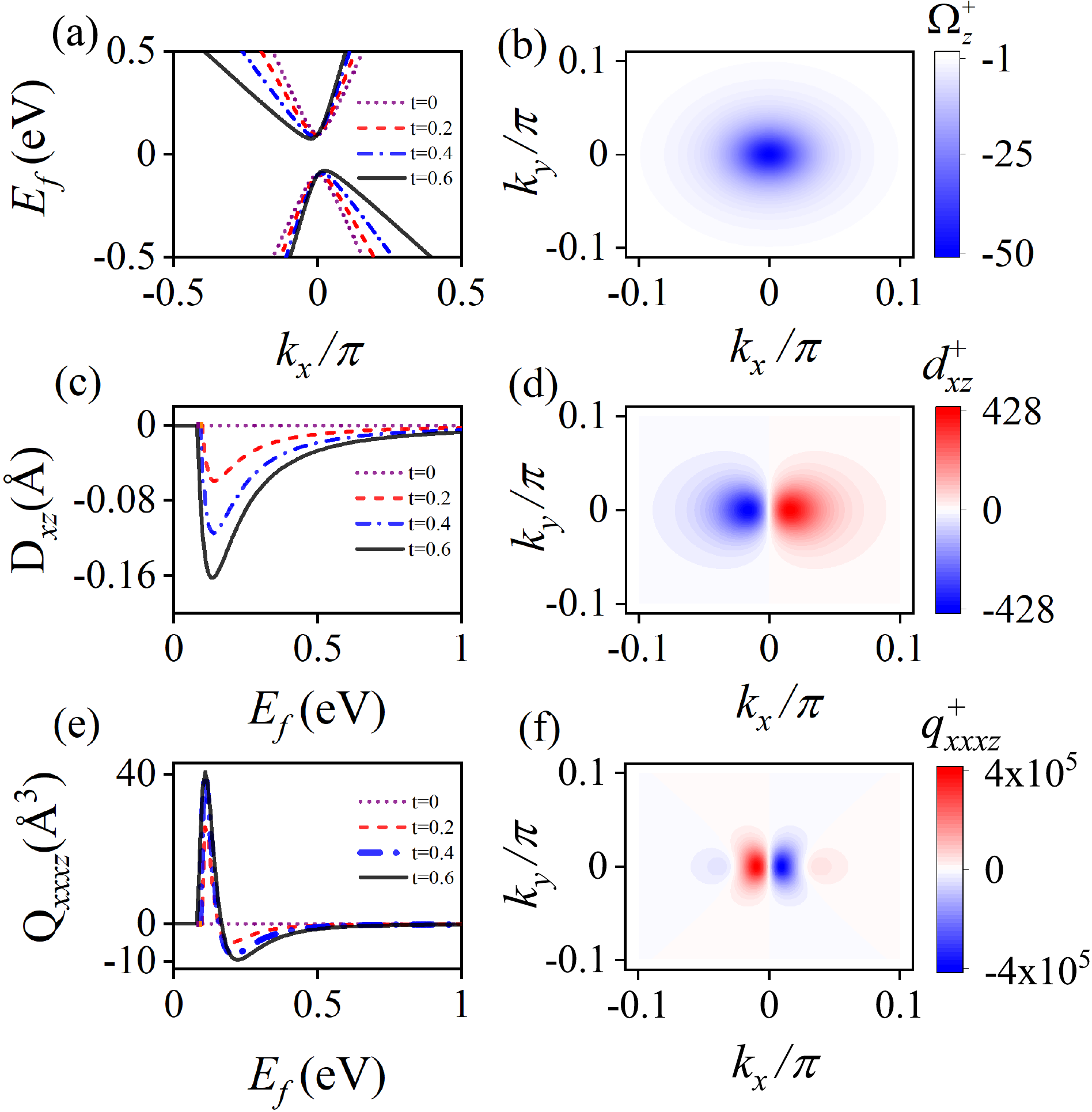}
\caption{(a) The energy structure of the tilted  2D  massive Dirac model in Eq. (\ref{infinite order13}) at $k_y=0$. (b), (d), and (f) The distribution of the Berry curvature $\Omega$, Berry dipole density $d_{xz}^+={\partial _x}\Omega _z^ +$, and Berry quadrupole density $ q_{xxxz}^+= {\partial^3 _x}\Omega _z^ +$ in momentum space.  The Berry curvature dipole and quadrupole vary with the Fermi energy $E_f$ in (c) and (e), respectively. All these are only for the conduction band. The other parameters are $v=1$ eV\AA, $t/v$ = 0, 0.2, 0.4, 0.6, and $\Delta$ = 0.1 eV.}
\label{fig1}
\end{figure}

\begin{figure}[tbp]
\centering
\includegraphics[width=1\linewidth]{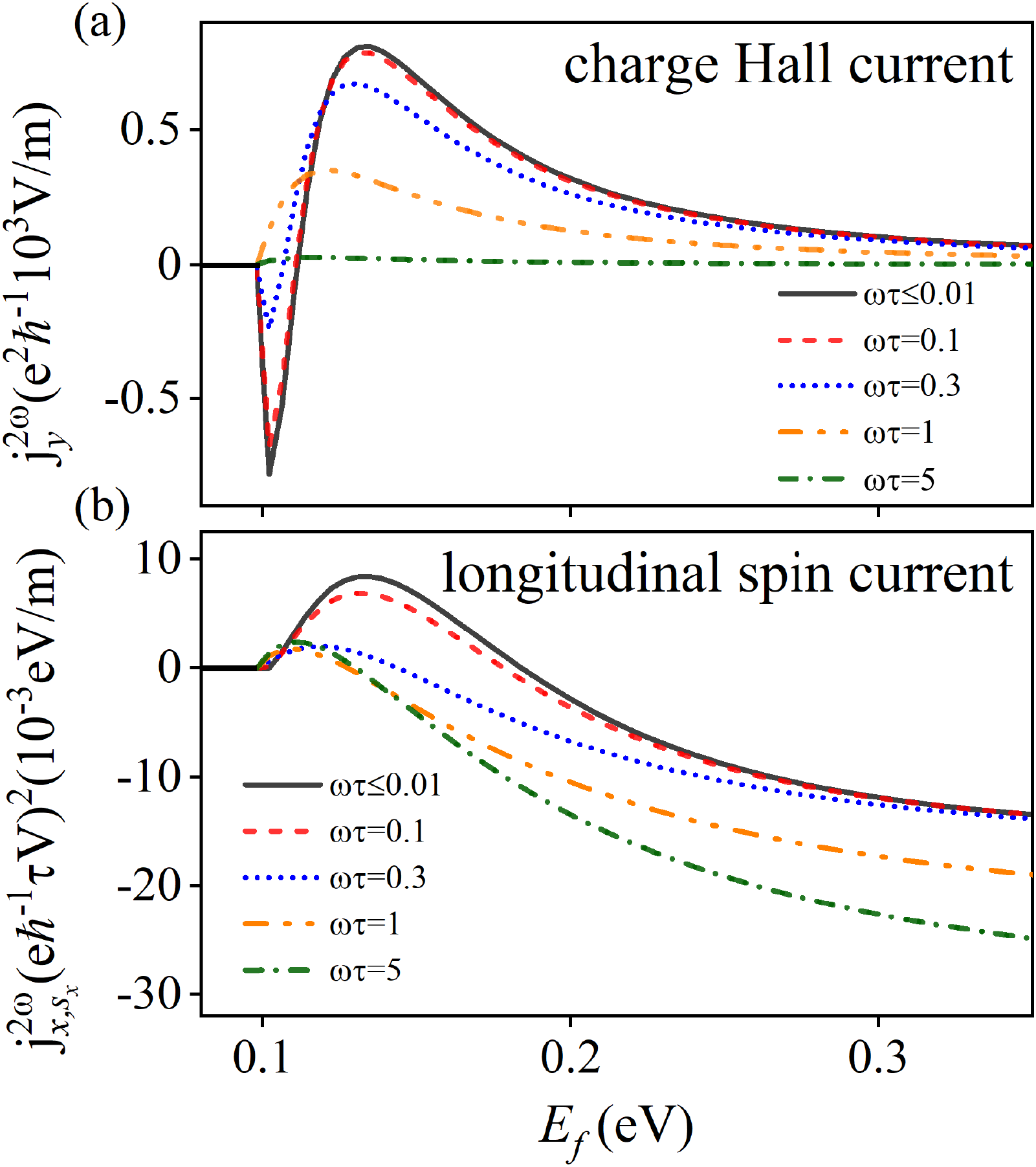}
\caption{(a) The charge current and (b) spin  current as a function of Fermi energy $E_f$. Here, the driving electric field is in the $x$ direction and the magnitude is ${\cal E}_x $ = $4 \times 10^5$ V/m, and $\tau$ = 1 ps. The other parameters are $v=1$ eV\AA, $t/v$ = 0.2, and $\Delta$ = 0.1 eV.}
\label{fig2}
\end{figure}

\emph{A tilted  2D  massive Dirac model.}---
Now we illustrate this general theory by studying a tilted  2D  massive Dirac model \cite{Du2019}. The Hamiltonian
 \begin{equation}
 \hat{\mathcal{H}}=t k_{x}+v\left(k_{x} \sigma_{x}+k_{y} \sigma_{y}\right)+\Delta \sigma_{z},
 \label{infinite order13}
 \end{equation}
where $(k_x,k_y)$ are the wave vectors, $(\sigma _x,\sigma _y,\sigma _z)$ are the Pauli matrices, and $t, v$ and $\Delta$ are the model parameters. $t$ tilts the Dirac cone along the $x$ direction. The Berry curvature is $\Omega_{\mathbf{k}}^{\pm}=\mp \frac{\Delta v^{2}}{2\left(v^{2} k^{2}+\Delta^{2}\right)^{3 / 2}}$, and the energy dispersions is $E_{\mathbf{k}}^{\pm}=t k_{x} \pm E_{\mathbf{k}}^{0}, \quad E_{\mathbf{k}}^{0}=\sqrt{v^{2} k^{2}+\Delta^{2}}$.
We can obtain all even-order charge current and only explicitly up to the fourth harmonic order ${j_a} = {\rm{Re}}\{ j_a^{2\omega }{e^{i2\omega t}} + j_a^{4\omega }{e^{i4\omega t}} +  \cdots \}$, where
\begin{eqnarray}
j_a^{2\omega }& =&  - \frac{{{e^3}\tau }}{{2{\hbar ^2}{Z_1}}}{\varepsilon ^{abc}}{D_{db}}{{\cal E}_c}{{\cal E}_d} - \frac{{{e^5}{\tau ^3}}}{{8{\hbar ^4}{C_1}}}{\varepsilon ^{abc}}{Q_{degb}}{{\cal E}_c}{{\cal E}_d}{{\cal E}_e}{{\cal E}_g}, \notag\\
j_a^{4\omega } &= & - \frac{{{e^5}{\tau ^3}}}{{8{\hbar ^4}{Z_3}}}{\varepsilon ^{abc}}{Q_{degb}}{{\cal E}_c}{{\cal E}_d}{{\cal E}_e}{{\cal E}_g}, \label{infinite order14} \\
{D_{db}}& =& \sum\limits_I {\int_k {\frac{{\partial \Omega _b^I}}{{\partial {k_d}}}f_0^I} } ,\quad {Q_{degb}} = \sum\limits_I {\int_k {\frac{{{\partial ^3}\Omega _b^I}}{{\partial {k_d}\partial {k_e}\partial {k_g}}}} } f_0^I, \notag
\end{eqnarray}
where $\cal E={\cal E}^*$, $D_{db}$ is the BCD \cite{Sodemann}, $d_{db}^I = {\frac{{\partial \Omega _b^I}}{{\partial {k_d}}}}$ is the Berry dipole density of the $I$-th band \cite{Du}, ${Z_n} =\prod\limits_{n'=1}^n {(1+i n' \omega \tau)}$,
$C_1(\omega ,\tau )$ is a dimensionless coefficient \cite{SM}.
 Similarly, we define a new quantity ${Q_{degb}}$ as the BCQ, and the
 $ q_{degb}^I = \frac{{{\partial ^3}\Omega _b^I}}{{\partial {k_d}\partial {k_e}\partial {k_g}}}$ as the Berry quadrupole density.
 It is instructive to study the case in a weak external electric field. For each $2m$-th harmonic order charge current density, it generally contains infinite terms of expansion order of the electric field charge current density whose order is larger than and equal to $2m$. However, after neglecting the higher expansion order of the electric field charge current density, we only keep the $2m$-th expansion order of the electric field (the leading order in this case), and have \cite{SM}
\begin{equation}
\begin{aligned}
j_y^{2m\omega} \propto {\cal E} _x^{2m}\sum\limits_I {\int_k {(\partial _x^{2m - 1}\Omega _z^I){f_0^I}} }.
\label{infinite order15}
\end{aligned}
\end{equation}
It says that any $2m$-th (even) harmonic order current density is mainly contributed from $2m$-th expansion order of the electric field   charge current density, which is induced by a $2m$-th multipole moment of Berry curvature (the integration). The higher multipole moment of Berry curvature is introduced in this work and emerges naturally. For a weak driving electric field, this multipole moment of Berry curvature can be explored in principle by measuring the higher harmonic order response charge current in experiments.

When the driving electric field is in the $x$ direction and only a single mirror symmetry exists, the BCD and BCQ are forced to be orthogonal to the mirror line \cite{Sodemann,Du}. Meanwhile, the charge current in Eq. (\ref{infinite order14})  can be expressed in vector notations as
\begin{eqnarray}
\vec{j}^{2\omega}&=& - \frac{{{e^3}\tau }}{{2{\hbar ^2}{Z_1}}}\hat{z} \times  \vec{{\cal E}} ({\vec{D}} \cdot  \vec{{\cal E}}) - \frac{{{e^5}{\tau ^3}}}{{8{\hbar^4}{C_1}}} \hat{z} \times  \vec{{\cal E}}(\vec{Q} \cdot \vec{{\cal E}})(\vec{{\cal E}}\cdot  \vec{{\cal E}}),\notag\\
\vec{j}^{4\omega } &= &- \frac{{{e^5}{\tau ^3}}}{{8{\hbar ^4}{Z_3}}}\hat z \times \vec {\cal E} (\vec Q \cdot \vec{\cal E} )(\vec {\cal E}  \cdot \vec{\cal E} ).
\label{infinite order16}
\end{eqnarray}

In addition to the band dispersion, Berry curvature, BCD, BCQ and their density distributions in momentum space are shown in Fig. \ref{fig1}.
If there is no titling ($t=0$), the BCD \cite{Du} and BCQ are both zero [see Figs. \ref{fig1}(c) and (e)], which is due to the symmetric band [ Fig. \ref{fig1}(a)]. For nonzero titling, a negative peak of BCD [ Fig. \ref{fig1}(c)] appears near the conduction band edge ($E_f \approx 0.12$ eV), while the BCQ [Fig. \ref{fig1}(e)] displays a positive peak ($E_f \approx 0.10$ eV) followed by a negative one ($E_f \approx 0.21$ eV). The different features are mainly caused by different distributions of the BCD and BCQ densities in $k$ space [Figs. \ref{fig1}(d) and (f)]. Although both have the symmetry of $C_2$, the BCD has one node at the center of the Dirac cone ($k=0$) as it contains only the first derivative of Berry curvature to $k_{x}$. Analogously, the BCQ develops  three nodes which are distributed symmetrically around $k=0$ since it contains the third derivative of Berry curvature. We emphasize that the BCD and BCQ characterized by $\vec{D}$ and $\vec{Q}$ in a vector form in Eq. (\ref{infinite order16}) can be directly determined by measuring the second and fourth harmonic order current density in experiments.


Figure \ref{fig2} shows the Fermi energy $E_f$ dependence of the 2HCH current density in Fig. \ref{fig2}(a) and the second harmonic order  longitudinal spin current density in Fig. \ref{fig2}(b). For small $\omega\tau$ (slow field, dirty limit), the 2HCH current density [Fig. \ref{fig2}(a)] develops a negative and a positive peak with increasing $E_f$. The negative peak is due to the effect of BCQ around the bottom of the conduction band. With increasing $\omega\tau$ to larger than 1 (fast field, clean limit), the negative peak gradually shrinks and disappears eventually, indicating a weakening effect of BCQ for larger $\omega\tau$. Another feature is that the 2HCH current tends to be independent of $\omega$ for very small  $\omega\tau$.
For the spin current, the model shows that there is only longitudinal even harmonic order spin current density and the transverse even harmonic order, and longitudinal and transverse odd harmonic order spin current densities are all zero. The reason is that the broken inversion symmetry due to tilting in the $x$ direction only releases the spin degeneracy in this direction, leading to the nonvanishing nonlinear longitudinal spin current (even harmonic order). Figure \ref{fig2}(b) only shows the nonvanishing second harmonic order  longitudinal spin current densities varying from a zero value ($E_{f}$ in the band gap) to a positive peak ($E_{f}$ in conduction band). With further increasing $E_{f}$, it turns negative and gradually saturates. The saturation is due to a full effect of spin degeneracy breaking. Increasing $\omega\tau$ leads to an enhancement of this positive to negative transition.

\emph{Conclusions and Discussions.}---We develop a theory of nonlinear responses driven by an ac electric field for charge and spin transport in the presence of nonvanishing Berry curvature. We introduce the harmonic order  and the expansion order of the electric field  and find that every harmonic order  current is composed of an infinite expansion order of the electric field, and the lowest expansion order of the electric field is equal to the harmonic order. Therefore, the nonlinear charge Hall effect studied previously just results from the second harmonic order and second expansion order of the electric field. To characterize the effects of a higher expansion order of the electric field, we introduce the Berry curvature multipole moment in which the so-called Berry curvature dipole is derived from the second expansion order of the electric field. We identify a ``selection rule" for the nonlinear charge and spin currents for four cases with respect to the TRS and IS. Finally, we show the role of the Berry curvature quadrupole moment by studying a specified model. A vector formalism of the charge current is given, indicating the Berry curvature quadrupole may be inferred in experiments via this equation. In real applications, the order of harmonics may not be very large in some cases. For a material with an energy gap and the Fermi level is lying in the middle of the gap, a driving harmonics with high enough order may induce interband transitions, which is beyond the scope of our paper. Another case which deserves to mention is that when the energy bands are higher above the Fermi level, a resonant interband transition could still be happening for a large enough $n$. As our theory may be more important in slow fields and in dirty doping cases, there is plenty of space to test the predicted effect from higher harmonics in realistic applications.

This work is supported in part by the National Key R\&D Program of China (Grant No. 2018YFA0305800), and the NSFC (Grants No. 11974348 and No.11834014). It is also supported by the Fundamental Research Funds for the Central Universities, and the Strategic Priority Research Program of CAS (Grants No. XDB28000000, and No. XDB33000000).

%

\end{document}